# FLUX EXPULSION AND MATERIAL PROPERTIES OF NIOBIUM EXPLORED IN 644-650 MHZ CAVITIES

K. McGee, S-H. Kim, P. Ostroumov, FRIB/MSU, East Lansing, MI 48824, U.S.A.
G. Eremeev, F. Furuta, M. Martinello, O. Melnychuk, A. Netepenko, Fermilab, Batavia, IL 60510

*Abstract*

Upcoming projects requiring high-Q ~650 MHz medium-to-high-$\beta$ elliptical cavities drive a need to understand magnetic RF loss mechanisms and mitigations in greater detail. High-temperature annealing and fast-cooldowns have proven effective techniques for promoting magnetic flux expulsion in cavities, however the extent of their effectiveness has been observed to vary between niobium material lot and vendor. We explore the fast-cooldown method, and high-temperature annealing (900°C) in 644-650 MHz cavities fabricated from two different niobium vendors: Tokyo-Denkai, and Ningxia. which promote flux-expulsion efficiency. Using EBSD and PPMS methods, we aim to trace cavity flux expulsion efficiency to specific, measurable properties of the bulk niobium material, which, if identified, can lead to methods by which the flux expulsion properties of Nb material can be predicted prior to cavity fabrication, and can enable fine-tuning of cavity temperature treatments.

## INTRODUCTION

Two upcoming projects, the Proton Improvement Plan II (PIP-II) at Fermilab [1] and the development work for Michigan State University's Facility for Rare Isotope Beams 400 MeV/u Uranium upgrade (FRIB400) [2] employ $\beta$ = 0.6 650 and 644 MHz 5-cell elliptical superconducting RF cavities to meet their project design goals. Both machines require continuous wave (CW)-compatible cavities, driving the cavity quality factor ($Q_0$) requirements into the realm of greater than 2 x $10^{10}$ at operating gradients of 16.8-17.5 MV/m [1,3].

The European Spallation Source (ESS) and the Spallation Neutron Source (SNS) use cavities of a similar size and frequency, however both of these machines operate in pulsed mode, and thus have less stringent requirements for $Q_0$—nearly an order of magnitude lower than that of PIP-II or FRIB400. Buffered chemical polishing (BCP) was employed to meet the $Q_0$ requirements of these machines, however, it has been demonstrated this treatment is unlikely to meet the needs of CW machines such as PIP-II or FRIB400 [4]. While [4] showed electropolishing (EP) is capable of meeting the minimum $Q_0$ design goal of, for example, FRIB400, a higher margin in $Q_0$ is desirable for a number of reasons: increasing $Q_0$ by even a few factors would ensure the FRIB400 upgrade fit within a modest upgrade of existing cryogenic facilities, and provide substantial operational cost savings. Moreover, exceeding the minimum $Q_0$ design goal by higher margins would guard against the possibility of $Q_0$ dropping below the minimum as a result of the cavity performance degradation that is sometimes observed between the cavity vertical test results and ultimate cryomodule results.

Nitrogen-doping (N-doping) and furnace baking are two advanced high-$Q_0$ treatments that have shown significant ability to improve $Q_0$ in 1.3 GHz cavities, and preliminary studies show they also have the ability to benefit 644-650 MHz cavities, with some refinements in method [5,6]. While the high-$Q_0$ potential of these recipes highly motivates their application in these cavities, a well-documented drawback of these methods is their propensity to increase the niobium cavity's sensitivity to trapped magnetic flux, $S$ [7]; $S$ being a measure of how many n$\Omega$ of resistance is contributed to the cavity's surface resistance, $R_S$, per mG of trapped flux, via increases in the temperature-independent component of $R_S$, known as the residual resistance, $R_0$.

The fractional contributions of $R_0$ to $R_S$ in cavities with high $S$ values can severely limit cavity performance if measures are not taken to reduce or eliminate magnetic flux trapping in cavities. While the first measure against flux trapping is always to implement good magnetic shielding and practice good magnetic hygiene in cryomodule construction to eliminate the background magnetic flux available for trapping in the cavity environment, other techniques that promote efficient cavity flux expulsion are necessary in the event stray magnetic field is still present. High thermal gradients, measured as a temperature differential (dT) over a unit of space, (dx) were shown to be key to promoting good flux expulsion in cavities, and furnace treatments in the range of 900 C were found to enhance this effect [8].

However, in the course of broadening N-doping studies, it became apparent that cavities fabricated from different vendors' niobium, or even from different lots of niobium from within the same vendor, could have varying degrees of flux expulsion efficiency. Further, it was observed that these responded differently to heat treatment: a heat treatment of a certain duration and temperature could completely rescue the flux-expulsion performance of one cavity whilst having minimal effect on another. Since it was clear from [8] that flux-trapping was a bulk property of the niobium, the question followed, what features of the material functioned as flux pinning centres, and what effect was the heat treatment having on them.

Identifying measurable material parameters with the potential to predict ultimate cavity flux expulsion performance raises the possibility that future material specifications could be written for niobium vendors to ensure laboratories could purchase lots of niobium with good flux expulsion properties. Moreover, thermal treatments could then be explored in small Nb samples, instead of after prototype cavity fabrication.

## CAVITY FLUX EXPULSION MEASUREMENTS

In the cryomodule and vertical testing contexts, some amount of Earth's background magnetic field is generally present. With shielding and reasonable magnetic hygiene practices, this field can be constrained to a few mG or so:

specification for within the FRIB baseline cryomodules, consisting of buffered chemical polished cavities, is for an environment of 15 mG or less, and PIP-II, anticipating the possible use of flux-sensitive cavity treatments such as N-doping, has written specifications for 5 mG or less.

For this study, cavity flux expulsion was quantified as a ratio between the magnetic field measured at the cavity equator before the superconducting transition ($B_{nc}$) and after the superconducting transition ($B_{sc}$).

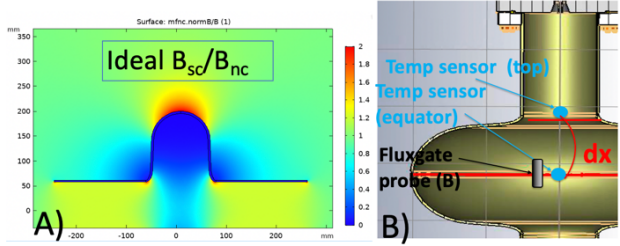

Figure 1: A) Ideal flux expulsion ($B_{sc}/B_{nc}$) simulated for a 650 MHz niobium cavity, showing ideal ratio of ~1.8 near the equator. B) Diagram of the experimental measurement of $B_{sc}/B_{nc}$. The thermal gradient (dT/dx) is measured as the difference in temperature between the two measurement points when the equator reaches Tc, where dx is the linear distance between the two temperature sensors.

The ideal flux expulsion ratio ($B_{sc}/B_{nc}$) can be found by means of simulation. Figure 1 A shows a COMSOL simulation of "perfect" flux expulsion in a $\beta_{opt}$ = 0.6 cavity, demonstrating that a measured ratio of around ~1.8 corresponds to a cavity that expels field well. Measured ratios of below 1.8 thus indicate some amount of flux trapping.

Flux expulsion is usually measured as a function of spatial thermal gradient (dT/dx), and has been demonstrated to increase as a function of (dT/dx), which motivates the "fast-cooldown" method of maximising flux expulsion. The method by which dT/dx is measured is shown in Figure 1 B: two thermal sensors are mounted on the cavity at the equator and the iris, and the difference in temperature (dT) between the two is measured when the thermal sensor at the equator reaches Tc (9.2 K). dx is taken as the linear surface distance between the two sensors (in cm). While this measure is comparable between cavities of the same size and shape, it is worth noting that when comparing across different cavity shapes, it does not capture the difference in area of the superconducting transition region as it traverses the cavity.

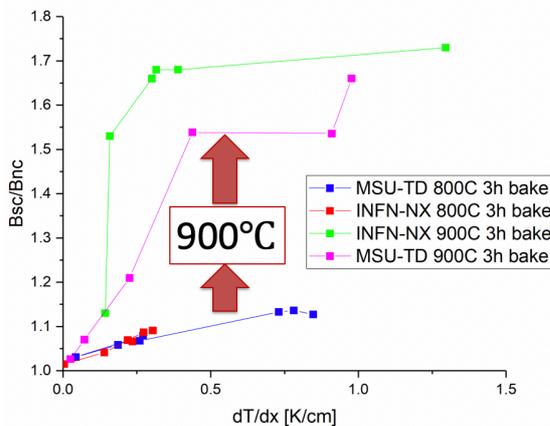

Figure 2: Flux expulsion ratio (y-axis) vs cooldown spatial gradient (dT/dx) of $\beta_{opt}$ = 0.6 650 MHz cavities fabricated from Ningxia (red/green) and Tokyo-Denkai (blue/magenta) material.

Figure 2 shows the results of flux expulsion measurements of two $\beta_{opt}$ = 0.6 ~650 MHz cavities, one fabricated from Tokyo-Denkai material (MSU-TD) and one fabricated from Ningxia material (INFN-NX). Initially, after just 800C hydrogen degassing, both cavities perform relatively poorly. After 900C baking, $B_{sc}/B_{nc}$ improves significantly for both cavities.

## GRAIN BOUNDARIES

Initial hypothesis [8] suggested that the grain boundaries were the primary pinning centres for the magnetic flux lines, thus high-temperature annealing ameliorated flux expulsion by increasing grain size and reducing grain boundaries.

Scrap material from both cavities' fabrication was recovered, and samples were cut and prepared for EBSD grain size measurements. Note, since the material was cut from the flat scrap material, it was neither worked or drawn in the way that the corresponding cavity material has been. These samples were cut and mounted in the cross-sectional direction, and the grains imaged were from the middle region of this cross-section.

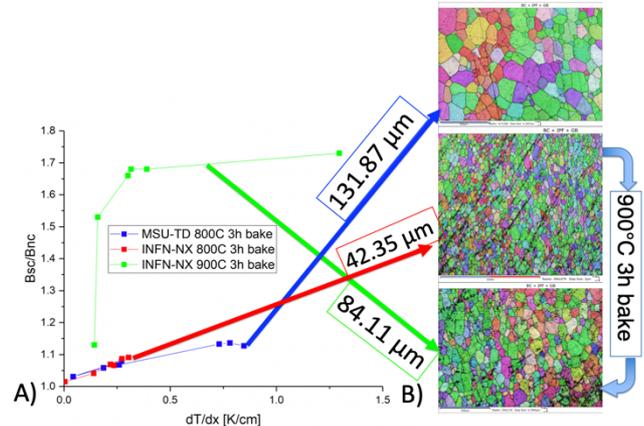

Figure 3: A) flux expulsion performance vs. cooldown gradient. 900°C 3 h baking significantly improved Bsc/Bnc of the Ningxia (NX) material. B) EBSD images of corresponding scrap Nb samples from cavity fabrication. Top: Tokyo-Denkai (TD) 800°C 3 h bake, Middle: NX 800°C 3 h bake, Bottom: NX 900°C 3 h bake.

Figure 3 draws the comparison between the results of EBSD grain size imaging and the corresponding cavities' flux expulsion performance. Despite a difference in average elliptical grain size of around 90 microns, both MSU-TD and INFN-NX material performed similarly before 900C baking. 900C baking improved the INFN-NX material, and roughly doubled the grain size from the initial state. However, this grain size, corresponding to the best-expelling cavity, is still nearly 50 microns smaller than the poorly-performing MSU-TD material, clearly demonstrating that grain boundaries do not function as the primary pinning centres. Concurrent work [10,11] proposes the distribution of dislocations, which are associated with coldworking the material in addition to the grain boundaries, are the more significant flux pinning centres, and are what are dissolved with high-temperature annealing.

## FLUX PINNING FORCE

The results of the previous section indicate that grain size cannot be used as a predictor of how well a cavity fabricated from that material will expel flux. The Fermilab Material

Science Laboratory (MSL) has a Physical Property Measurement System (PPMS) available, which enables direct measurement of the flux-pinning force acting on a small 3 mm x 8 mm cylindrical sample of Nb. This is, in essence, direct measurement of the force that must be overcome by the force generated by the thermal gradient in order to expel the magnetic flux from the Nb bulk.

This measurement was conducted by cooling the sample to 9 K, which is the closest temperature to the critical temperature of niobium (9.2 K) that can be stably maintained in this apparatus. Magnetic field is then imposed on the sample, in alternating polarizations, and the resultant irreversible magnetization of the sample is measured. From the irreversible magnetization, the associated circulating current around the pinned flux lines can be derived, which then enables the force to be calculated from the applied magnetic field.

The magnetic field in this measurement is on the order of Gauss, which is obviously not representative of the few mGauss expected in the test Dewar or cryomodule. The aim of measuring the flux-pinning force is instead to find some fundamental quantity that correlates with the trend noted in cavity flux expulsion performance, not to replicate the conditions of the test Dewar or cryomodule. The results are shown in Figure 4.

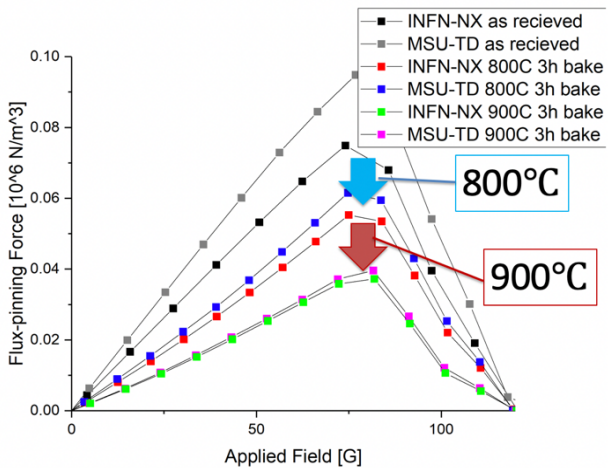

Figure 4: The measured flux-pinning force in the corresponding Nb samples. 900°C heat treatment improved the flux expulsion performance in both samples, which corresponds with a similar decrease in measured flux pinning force in both samples. This suggests flux-pinning force, and not grain size, is the better predictor of cavity flux expulsion performance.

A first reduction in flux-pinning force is seen after the MSU-TD and INFN-NX material samples undergo 800C baking, representative of the hydrogen degassing the cavities underwent. Subsequent 900C baking of the samples provided further reduction in the measured flux-pinning force, notably, quite similar reductions in both the NX and TD samples, despite the previously mentioned disparate grain sizes.

Somewhat unsurprisingly, thus, we show that the flux-pinning force measurement appears to probe the more fundamental property related to the ability of the cavity to expel magnetic flux. This is a promising correlation, but requires more statistical evidence before an absolute flux-pinning measurement can be predictively related to a specific flux expulsion ratio. However, if such an understanding can be established, PPMS measurements may prove a useful method by which good flux-expelling lots of Nb can be separated from poor flux-expelling lots of Nb, prior to cavity fabrication.

## IMPLICATIONS FOR Q

In order to underscore the importance of understanding and learning to manipulate the material parameters that contribute to flux expulsion properties in Nb, flux sensitivity measurements, ($S$, in n$\Omega$/mG) such as the one described in [6], have been used to produce the chart in Figure 6. Figure 6 depicts the expected reduction in $Q_0$, at the FRIB400 operating gradient of 17.5 MV/m, as a function of various amounts of trapped flux. The slope of the line of points corresponds to the measured flux sensitivity, with higher sensitivities corresponding to a steeper slope.

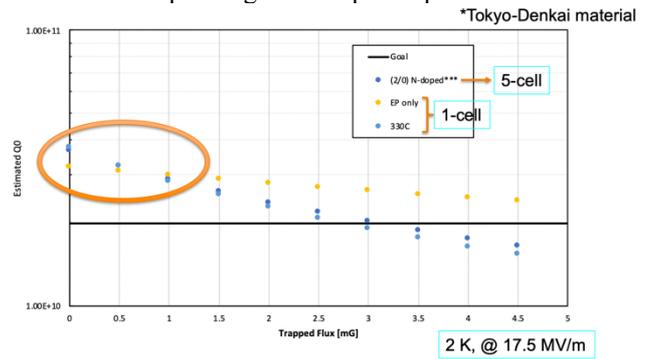

Figure 6: A) flux sensitivity experimental setup. B) projected Q based on measured sensitivities in 644 MHz cavities (1-cell and 5-cell as noted) at 17.5 MV/m for Tokyo-Denkai material.

The critical observation from this plot is that at around 1 mG of trapped flux or less, a certain regime dominates in which the N-doping or furnace baking (330C) treatments provide the higher $Q_0$. If 1 mG or more of trapped flux remains after cooldown, the EP treatment provides the higher $Q_0$. Understanding and manipulating the physical parameters previously mentioned can help drive the operational regime into the circled region, however, if this is not possible, a realistic approach must be taken regarding the choice of which RF surface treatment to choose for production cryomodules.

## CONCLUSIONS

While both TD and NX material benefit from high-temperature annealing, flux expulsion is not purely a function of grain size, and dislocation density appears to have the stronger effect on flux expulsion [10]. Flux pinning force more directly correlates with expulsion. While 900°C annealing improves flux expulsion, applying this treatment to large cavities may compromise mechanical properties, making improved shielding the more practicable solution.

## ACKNOWLEDGEMENTS

This document was prepared in conjunction with MSU/FRIB using the resources of the Fermi National Accelerator Laboratory (Fermilab), a U.S. Department of Energy, Office of Science, HEP User Facility. Fermilab is managed by Fermi Research Alliance, LLC (FRA), acting under Contract No. DE-AC02-07CH11359.

# REFERENCES

[1] M. Ball, A. et al., "The PIP-II Conceptual Design Report." White paper. 2017. https://lss.fnal.gov/archive/design/fermilab-design-2017-01.pdf

[2] FRIB400 White Paper; https://indico.frib.msu.edu/event/2/attachments/57/292/FRIB400_Final.pdf.

[3] P. N. Ostroumov et al., "Elliptical superconducting RF Cavities for FRIB energy upgrade." in *Nuclear Inst. and Methods in Phys. Research,* A. **888**, 53-63, 2018.

[4] K. McGee, S. Kim, K. Elliott, A. Ganshyn, W. Hartung, E. Metzgar, P. N. Ostroumov, et al., "Medium-Velocity Superconducting Cavity for High Accelerating Gradient Continuous-Wave Hadron Linear Accelerators." Phys. Rev. Accel. Beams, **24**, 112003 (2021).

[5] M. Martinello et al. "Q-factor optimization for high-beta 650 MHz cavities for PIP-II", J. Appl. Phys. **130** 174501 (2021).

[6] K. McGee, K. Elliott, A. Ganshyn, W. Hartung, S. Kim, P. Ostroumov, et al., "Development towards FRIB upgrade to 400 MeV/u for the heaviest uranium atoms." In Proc. LINAC22, Liverpool, UK. TH1AA02. (2022).

[7] M. Checchin *et al*., "Electron mean free path dependence of the vortex surface impedance." Supercond. Sci. Technol. **30**, 034003 (2017)

[8] S. Posen *et al.,* "Efficient expulsion of magnetic flux in superconducting radio frequency cavities for high $Q_0$ applications." J. Appl. Phys. **119**, 213903 (2016).

[9] M. Martinello, "The Path to High Q-factors in Superconducting Accelerating Cavities: Flux Expulsion and Surface Resistance Opimization" Graduate Thesis, Ilinois institute of Technology, USA. 2017.

[10] Z.L. Thune *et al.,* IEEE Transactions on Appl. Supercond. **33**, 5 (2023).

[11] M. Martinello, A. Grassellino, S. Posen, A. Romanenko, Z. Sung, J. Lee. "Microscopic Investigation of Flux Trapping Sites in Niobium SRF Cavities" Poster. SRF'19 Dresden. THP013.